\documentstyle[preprint,tighten,aps,psfig]{revtex}
\begin{document}
\draft
\title{Electroweak corrections to the muon anomalous magnetic moment}
\author{Andrzej Czarnecki and Bernd Krause}
\address{Institut f\"ur Theoretische Teilchenphysik, \\
Universit\"at Karlsruhe,
D-76128 Karlsruhe, Germany}
\author{William J.~Marciano}
\address{Physics Department,
Brookhaven National Laboratory\\  Upton, New York 11973}
\maketitle
\begin{abstract}
The bosonic two-loop electroweak radiative corrections to the muon's
anomalous magnetic moment, $a_\mu\equiv (g_\mu-2)/2$, are presented.
We find $\Delta a_\mu^{\rm EW}({\rm 2\,loop\, bosonic})/
a_\mu^{\rm EW}({\rm 1\,loop})\approx   {\alpha\over \pi}\left(-3.6
\ln\left({M_W^2\over m_\mu^2}\right) +0.10 \right)\approx -0.11$ for
$M_{\rm Higgs}\approx 250$ GeV.  Combining that result with our
previous two-loop fermionic calculation, we obtain an overall 22.6\%
reduction in $a_\mu^{\rm EW}$ from $195\times 10^{-11}$ to 
$151(4)\times 10^{-11}$.  Implications for the full standard model
prediction and
an upcoming high precision measurement of $a_\mu$ are briefly discussed.
We also give the two-loop electroweak corrections to the anomalous
magnetic moments of electron and tau lepton; they result in a reduction
of the one-loop estimates by 35\% and 15\%, respectively.
\end{abstract}

The anomalous magnetic moment of the muon, $a_\mu\equiv (g_\mu-2)/2$,
provides both a sensitive quantum loop test of the standard
$SU(3)_C\times SU(2)_L\times U(1)$ model and a window to potential
``new physics'' effects.  The current experimental average \cite{PDG}
\begin{eqnarray}
a_\mu^{\rm exp} = 116592300 (840) \times 10^{-11}
\end{eqnarray}
is in good agreement with theoretical expectations and already
constrains physics beyond the standard model such as supersymmetry and
supergravity
\cite{km90,Nath95}, dynamical or loop muon mass generation \cite{3},
compositeness \cite{4}, leptoquarks \cite{Couture95} etc.

An upcoming experiment E821
\cite{Hughes92} at Brookhaven National Laboratory is
expected to start in 1996.  With one month
of dedicated running, it is expected to reduce the uncertainty 
in $a_\mu^{\rm exp}$ to roughly  $\pm 40\times 10^{-11}$, 
more than a factor of 20 improvement.
With subsequent longer dedicated runs it could statistically 
approach the anticipated
systematic uncertainty of about $\pm 10-20\times 10^{-11}$ \cite{Bunce}.
At those levels, both electroweak
one and two loop effects become important and ``new physics'' at the
multi-TeV scale is probed.  Indeed, generic muon mass generating
mechanisms (via perturbative or dynamical loops \cite{3}) lead to
$\Delta a_\mu \approx m_\mu^2/\Lambda^2$, where $\Lambda$ is the scale
of ``new physics''.  At $\pm 40\times 10^{-11}$ sensitivity, $\Lambda
\approx 5$ TeV is being explored.

To fully exploit the anticipated experimental improvement, the
standard model prediction for $a_\mu$ must be known with comparable
precision.  That requires detailed studies of very high order QED
loops, hadronic effects, and electroweak contributions through two
loop order.  The contributions to $a_\mu$ are traditionally divided
into 
\begin{eqnarray}
a_\mu=a_\mu^{\rm QED}+a_\mu^{\rm Hadronic}+a_\mu^{\rm EW}
\end{eqnarray}
QED loops have been computed to very high order \cite{r4,Kinoshita95}
\begin{eqnarray}
a_\mu^{\rm QED} &=& {\alpha\over 2 \pi}
\nonumber\\ &&
+0.765857381(51) \left( {\alpha\over  \pi}\right)^2
+24.050531(40) \left( {\alpha\over  \pi}\right)^3
\nonumber\\ &&
+ 126.02(42) \left( {\alpha\over  \pi}\right)^4
+ 930(170) \left( {\alpha\over  \pi}\right)^5
\label{eq:qed}
\end{eqnarray}
Employing $\alpha= 1/137.03599944(57)$ obtained from the electron
$g_e-2$, implies \cite{Kinoshita95}
\begin{eqnarray}
a_\mu^{\rm QED} &=& 116584706(2)\times 10^{-11}
\end{eqnarray}
The uncertainty is well within the $\pm 20-40\times 10^{-11}$ goal.
Indeed, even if we take the last known term in (\ref{eq:qed}) as
indicative of its truncation uncertainty, the QED error remains
relatively small.

Hadronic vacuum polarization corrections to $a_\mu$ enter at ${\cal
O}(\alpha/\pi)^2$.  They can be evaluated via a dispersion relation
using $e^+e^- \to hadrons$ data and perturbative QCD (for the
very high energy regime).  Employing a recent analysis of $e^+e^-$
data \cite{Jeg95} along with an estimate of the leading ${\cal
O}(\alpha/\pi)^3$ effects, we find  \cite{CKM95}
\begin{eqnarray}
a_\mu^{\rm Hadronic}({\rm vac.\,pol.})= 6934(153)\times 10^{-11}
\label{eq:hadr}
\end{eqnarray}
Unfortunately, the error has not yet reached the desired level of
precision.  Ongoing improvements in $e^+e^- \to hadrons$ 
measurements at low energies along with additional theoretical input
should significantly lower the uncertainty
in (\ref{eq:hadr}). Nevertheless, reducing the hadronic error below
$\pm 20 \times 10^{-11}$ or even $\pm 40 \times 10^{-11}$ 
remains a formidable challenge.

The result in (\ref{eq:hadr}) must be supplemented by hadronic light
by light amplitudes (which are of three loop origin)
\cite{light,haya95,Bijnens95}.  Here, we employ
a recently updated 
study by Hayakawa, Kinoshita, and Sanda \cite{haya95} which
gives
\begin{eqnarray}
a_\mu^{\rm Hadronic}({\rm light\, by \, light})= -52(18)\times
10^{-11}
\label{eq:light}
\end{eqnarray}
However, we note that the result is somewhat dependent on the low
energy model of hadronic physics employed and  continues to be
scrutinized.  Combining (\ref{eq:hadr}) and (\ref{eq:light})
leads to the total hadronic contribution
\begin{eqnarray}
a_\mu^{\rm Hadronic}= 6882(154)\times 10^{-11}
\end{eqnarray}

Now we come to the electroweak contributions to $a_\mu$, the main
focus of our work and the impetus for forthcoming experimental effort.
At the one loop level, the standard model predicts 
\cite{fls72,Jackiw72,ACM72,Bars72,Bardeen72}
\begin{eqnarray}
\lefteqn{a_\mu^{\rm EW}(\rm 1\,loop) =
{5\over 3}{G_\mu m_\mu^2\over 8\sqrt{2}\pi^2}}
\nonumber\\ && \times
\left[1+{1\over 5}(1-4s_W^2)^2
+ {\cal O}\left({m_\mu^2 \over M^2}\right) \right]
\nonumber \\
&& \approx 195 \times 10^{-11}
\label{eq:oneloop}
\end{eqnarray}
where $G_\mu = 1.16639(1) \times 10^{-5}$ GeV$^{-2}$, $M=M_W$ or
$M_{\rm Higgs}$, and the weak mixing angle
$\sin^2\theta_W\equiv s_W^2 = 1-M_W^2/M_Z^2=0.224$.  We can safely
neglect the ${\cal O}\left({m_\mu^2 / M^2}\right)$ terms in
(\ref{eq:oneloop}).

The one loop result in (\ref{eq:oneloop}) is about five to ten times the
anticipated experimental error.  Naively, one might expect higher order (2
loop) electroweak contributions to be of relative ${\cal O}(\alpha/
\pi)$ and hence negligible; however, that is not the case.  Kukhto,
Kuraev, Schiller, and Silagadze (KKSS)  \cite{KKSS} have shown that
some two loop electroweak contributions can be quite large and must be
included in any serious theoretical estimate of $a_\mu^{\rm EW}$ or
future confrontation with experiment.  Given the KKSS observation, a
detailed evaluation of the two loop electroweak contributions to
$a_\mu$ is clearly warranted.  Here, we report the complete results of
such an analysis.  

The two loop electroweak  contributions to $a_\mu^{\rm EW}$ naturally divide 
into so-called fermion and boson parts
\begin{eqnarray}
a_\mu^{\rm EW} = &&a_\mu^{\rm EW} ({\rm 1\,loop})
\nonumber\\
&&+a_\mu^{\rm EW}({\rm 2\,loop;\,ferm.})
+a_\mu^{\rm EW}({\rm 2\,loop;\,bos.})
\end{eqnarray}
The $a_\mu^{\rm EW}({\rm 2\,loop;\,ferm.})$ includes all two loop
electroweak corrections which contain closed fermion loops while all
other contributions are lumped into $a_\mu^{\rm EW}({\rm
2\,loop;\,bos.})$.  In a previous study \cite{CKM95}, we computed 
$a_\mu^{\rm EW}({\rm 2\,loop;\,ferm.})$. For $M_{\rm Higgs}
\approx$ 250 GeV it reduces $a_\mu^{\rm EW}$ by 11.8\%.    We have now
completed that effort by computing $a_\mu^{\rm EW}({\rm
2\,loop;\,bos.})$.  
Our results are described below.

The one-loop diagrams which contribute to the lowest order electroweak
corrections to $a_\mu$ are shown in fig.~1 (there is another diagram
obtained by exchanging $W$ and $G$ in 1(c) but its value is the same
as 1(c).  This is true also for mirror reflections of two loop
diagrams, and hence we do not depict them.)  The diagrams of fig.~1,
minus Schwinger's photon exchange diagram in 1(a),
lead to the formula (\ref{eq:oneloop}).

The two-loop diagrams fall in two general categories.  The first and
largest group consists of all diagrams which can be viewed as
corrections to the one-loop diagrams in fig.~1.  Those are one-loop
insertions in the propagators and vertices, but also non-planar
diagrams and diagrams with quartic couplings.  The second group
includes all the new types which appear at the two-loop level, as
shown in fig.~2.

  The complete set of all two-loop diagrams is quite
large, together with fermionic loops 
it includes the total of 1678 diagrams \cite{Nakazawa95}.
However, the diagrams with two or more scalar couplings to the muon
line are suppressed by an extra factor of $m_\mu^2/M_W^2$ and can be
discarded.  This is true already at the one-loop level, where one
neglects the diagrams with the Higgs boson loop and with two Goldstone
boson couplings to the muon. 
Making this approximation and taking advantage of the
mirror symmetry mentioned above reduces the number of relevant
diagrams to about 240 in the linear 't~Hooft-Feynman gauge.  This
number can be almost halved by choosing a
non-linear gauge \cite{Fujikawa73} 
in which the $\gamma W^\pm G^\mp$ vertex vanishes.  
We performed the calculation in both gauges to have
a sensitive check of the accuracy of our procedures.
For both gauges, two loop divergences are canceled by
counterterm insertions in the one loop diagrams of fig.~1.

The smallness of the muon mass compared to the electroweak scales
allows us to employ the asymptotic expansion method \cite{Smi94}.  In
the present calculation we also assume that mass of the Higgs is large
compared to $M_{W,Z}$ and compute the first two terms in the expansion
in $M_{W,Z}^2/M_H^2$.  In diagrams where both $W$ and $Z$ bosons are
present we also expand in their relative mass difference.  This
corresponds to an expansion in $\sin^2\theta_W$ and we keep the first
four terms in this expansion.  This number of powers is also
sufficient to obtain an exact coefficient of the large logarithms $\ln
M_W^2/m_\mu^2$; these terms are generated by diagrams with either $Z$
or $W$ boson and hence their coefficient is a polynomial, rather than
a series, in $\sin^2\theta_W$.

The large logs have been considered by KKSS \cite{KKSS} in the
approximation $\sin^2\theta_W=1/4$.  We obtain a slightly different
coefficient even in this special case.  The difference between KKSS
and our calculation is that KKSS did not consider the diagram shown in
fig.~3, where a contribution to the leading log comes from a loop with
the Higgs boson.  There is no Higgs mass suppression in this diagram
because of the $HG^+ G^-$ coupling. Without this diagram the result is
gauge dependent.  For example, this diagram vanishes exactly in the
non-linear gauge we adopted 
in the cross-check, but not in the linear gauge.  It should,
therefore, have 
been included in the linear gauge calculation as in KKSS.

Altogether, we find for the two-loop electroweak corrections
\begin{eqnarray}
\lefteqn{a_\mu^{\rm EW}({\rm 2\,loop;\,bos.})
= 
{m_\mu^2\alpha G_\mu \over 8\sqrt{2} \pi^3}}\nonumber \\ 
&& \times
\left(
\sum_{i=-1}^2 
\left[ a_{2i} s_W^{2i} + {M_W^2\over M_H^2}b_{2i} s_W^{2i}\right]
+{\cal O}(s_W^6)
\right)
\end{eqnarray}
with 
\begin{eqnarray}
a_{-2} &=&{19\over 36} -{99\over 8}S_2
-{1\over 24} \ln{M_H^2\over M_W^2} 
\nonumber \\
a_{0} &=&-{859\over 18}+11{\pi\over \sqrt{3}} + {20\over 9}\pi^2 
+{393\over 8}S_2
\nonumber \\ &&
-{65\over 9} \ln{M_W^2\over m_\mu^2} 
+{31\over 72} \ln{M_H^2\over M_W^2} 
\nonumber \\
a_{2} &=&{165169\over 1080}-{385\over 6}{\pi\over \sqrt{3}} 
- {29\over 6}\pi^2  +{33\over 8}S_2
\nonumber \\ &&
+{92\over 9} \ln{M_W^2\over m_\mu^2} 
-{133\over 72} \ln{M_H^2\over M_W^2} 
\nonumber \\
a_{4} &=&-{195965\over 864}+{265\over 3}{\pi\over \sqrt{3}} 
+ {163\over 18}\pi^2  +{223\over 12}S_2
\nonumber \\ &&
-{184\over 9} \ln{M_W^2\over m_\mu^2} 
-{5\over 8} \ln{M_H^2\over M_W^2} 
\nonumber \\
b_{-2} &=&{155\over 192} + {3\over 8}\pi^2 -{9\over 8}S_2
+{3\over 2} \ln^2{M_H^2\over M_W^2} 
-{21\over 16} \ln{M_H^2\over M_W^2} 
\nonumber \\
b_{0} &=&{433\over 36} + {5\over 24}\pi^2 -{51\over 8}S_2
+{3\over 8} \ln^2{M_H^2\over M_W^2} 
+{9\over 4} \ln{M_H^2\over M_W^2} 
\nonumber \\
b_{2} &=&-{431\over 144} + {3\over 8}\pi^2 +{315\over 8}S_2
+{3\over 2} \ln^2{M_H^2\over M_W^2} 
\nonumber \\ &&
-{11\over 8} \ln{M_H^2\over M_W^2} 
\nonumber \\
b_{4} &=&{433\over 216} + {13\over 24}\pi^2 +{349\over 24}S_2
+{21\over 8} \ln^2{M_H^2\over M_W^2} 
\nonumber \\ &&
-{49\over 12} \ln{M_H^2\over M_W^2} 
\label{eq:result}
\end{eqnarray}

and
\begin{eqnarray}
S_2 \equiv {4\over 9\sqrt{3}}{\rm Cl}_2\left({\pi\over 3}\right)
= 0.2604341...
\end{eqnarray}
We have used the mass shell renormalization prescription
\cite{ONSHELL}.  Part of the two-loop bosonic corrections have been
absorbed into the lowest order result, by expressing one-loop
contributions in eq.~(\ref{eq:oneloop}) in terms of the muon decay
constant $G_\mu$.

Employing $\sin^2\theta_W = 0.224$ and $M_H = 250$ GeV in 
eq.~(\ref{eq:result},\ref{eq:oneloop}) gives
\begin{eqnarray}
{a_\mu^{\rm EW}({\rm 2\,loop\, bosonic})\over
a_\mu^{\rm EW}({\rm 1\,loop}) }
\approx
{\alpha\over \pi} \left( -3.6 \ln {M_W^2\over m_\mu^2} + 0.10\right)
\end{eqnarray}
which corresponds to a 11.0\% reduction.  For comparison, the partial
leading log calculation of KKSS gave $-{49\alpha\over
15\pi}\ln(m_Z^2/m_\mu^2)$, a 10.3\% reduction.  

Combining our new result and previous fermionic two-loop calculation
leads to a total reduction of $a_\mu^{\rm EW}$ by a factor
$(1-97\alpha/\pi) \approx 0.77$ and the new electroweak prediction 
\begin{eqnarray}
a_\mu^{\rm EW} = 151(4) \times 10^{-11}
\label{eq:newaEW}
\end{eqnarray}
The assigned error of $\pm 4\times 10^{-11}$ is due to uncertainties
in $M_H$ and quark two loop effects.  It also allows for possible
three loop (or higher) electroweak contributions.  In that regard, we
note that our calculation of the $\ln(M_W/m_\mu)$ coefficients can be
combined with a renormalization group analysis to sum up leading log
corrections of the form $\left({\alpha\over \pi}\ln (M_W/m_\mu)\right)^n$,
$n=1,2,\ldots$; that analysis will be given in a future publication.

With minor modifications, our results give also the two-loop
electroweak corrections to anomalous magnetic moments of other
leptons.  For the electron we find for the combined fermionic and
bosonic loops
\begin{equation}
{a_e^{\rm EW} (2\,{\rm loop})  \over a_e^{\rm EW} (1\,{\rm loop}) }
\approx -150{\alpha\over \pi}
\end{equation}
The two-loop corrections result in a 35\% reduction of the one-loop
prediction. 
 
For the tau lepton the total result of two-loop bosonic and fermionic
loops is
\begin{equation}
{a_\tau^{\rm EW} (2\,{\rm loop})  \over a_\tau^{\rm EW} (1\,{\rm loop}) }
\approx -65{\alpha\over \pi}
\end{equation}
The fermionic contribution has been computed assuming charm quark mass
equal approximately to the tau mass.
In the case of the $\tau$-lepton the two-loop corrections amount to a
15\% reduction of the one-loop result.

Our final result in (\ref{eq:newaEW}) along with the current best
estimates for $a_\mu^{\rm QED}$ and $a_\mu^{\rm hadronic}$
are illustrated in table I.  For comparison, the 1990 values are also
given \cite{km90}.  
Changes reflect the evolution and continuing scrutiny
of theoretical studies.  At present
\begin{eqnarray}
a_\mu^{\rm theory} = 116591739(154) \times 10^{-11}
\end{eqnarray}
with extremely small QED and EW uncertainties.  What remains is to
reduce the hadronic uncertainty by a factor of 4 (or more) via
improved $e^+e^-\to hadrons$ data and additional theoretical input.
Then, one can fully exploit the anticipated improvement in
$a_\mu^{\rm exp}$ from E821 at Brookhaven, a measurement we anxiously
await. 

\vspace{2cm}

\section*{Acknowledgement}
We thank Professor T.~Kinoshita for updating us on his work.
A.C.~and B.K.~would like to thank Professor W.~Hollik for many helpful
discussions and advice, and Professor J.H.~K\"uhn for encouragement
and support.  We acknowledge the hospitality of the
Institute for Nuclear Theory at the University of Washington and Aspen
Center for Physics, where parts of this work were completed.  This
research was supported by BMFT 056 KA 93P; by ``Graduiertenkolleg
Elementarteilchenphysik'' at the University of Karlsruhe; and by
U.S.~Department of Energy under contract number~DE-AC02-76CH00016.

\begin{table}
\caption{Update in $a_\mu \equiv { g_\mu -2\over 2}$ since the 1990
estimate.
All numbers have to be multiplied by $10^{-11}$.}
\vspace*{.1cm}
\begin{tabular}{lrrr}  
  & Current Value & 1990 Estimate & Change \\ \hline
  & & & \\
$ a_\mu^{QED} $ & $ 116\,584\,706\, \hspace*{.3cm}(2) $ & 
$ 116\,584\,696\,\hspace*{.3cm} (5)$ & $ +10$ \\  
$ a_\mu^{hadronic} $ & $ 6\,882\,(154) $ & $ 7\,027\, (175)$ & $ -145$ \\ 
$ a_\mu^{EW} $ & $ 151\,\hspace*{.3cm}(4) $ & $ 195\,\hspace*{.15cm}
(10)$ & $ -44$ \\   \hline
  & & & \\
$ a_\mu^{theory} $ & $ 116\,591\,739\, \hspace*{0cm}(154) $ & $
         116\,591\,918\, \hspace*{0cm}(176) $ & $ -179$   
\end{tabular}
\end{table}

\newpage
\begin{figure}
\begin{minipage}{25.cm}
\hspace*{.8cm}
\[
\mbox{
\hspace*{-67mm}
\begin{tabular}{ccc}
\psfig{figure=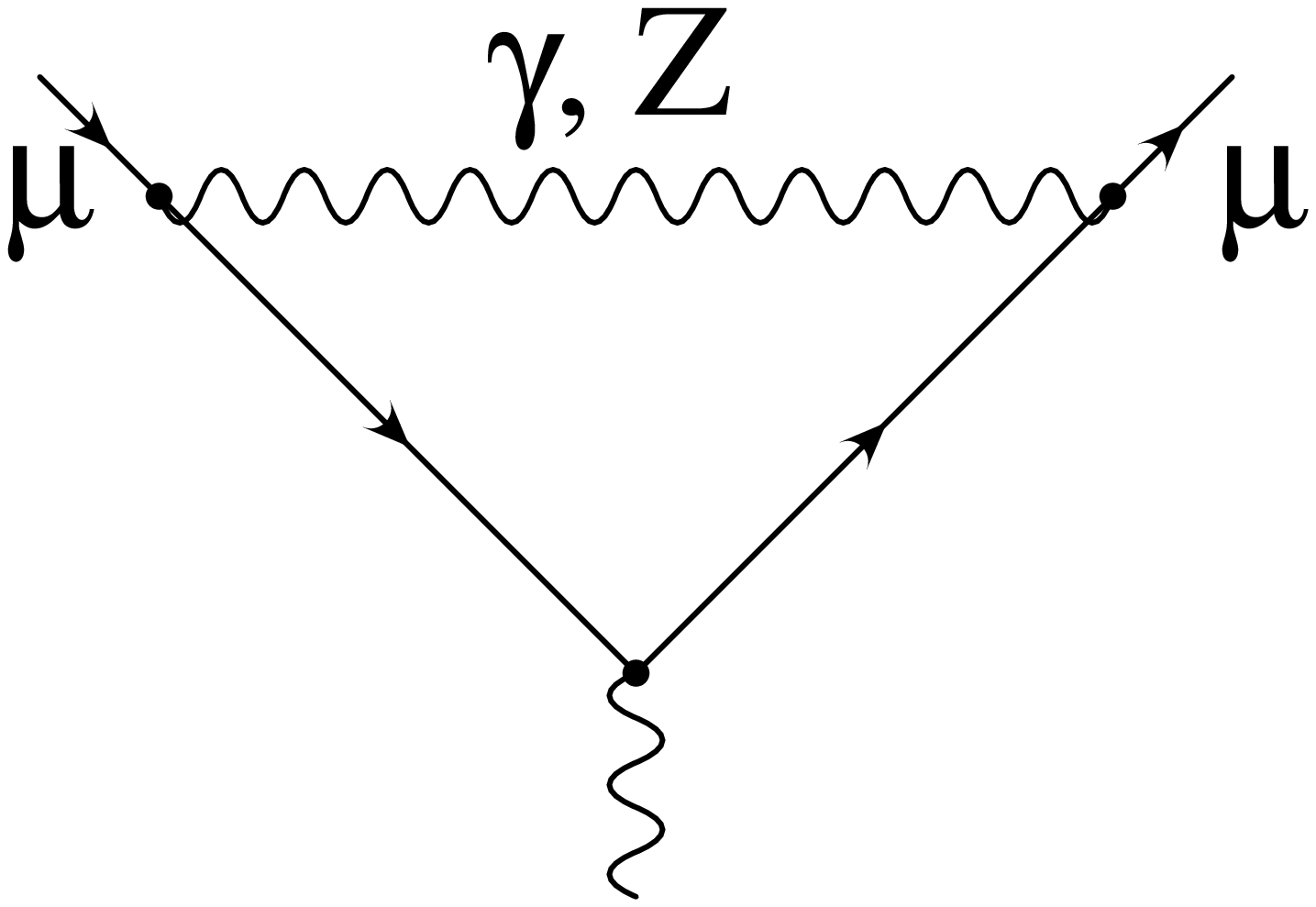,width=30mm,bbllx=210pt,bblly=410pt,bburx=630pt,bbury=550pt} 
&\hspace*{.6cm}
\psfig{figure=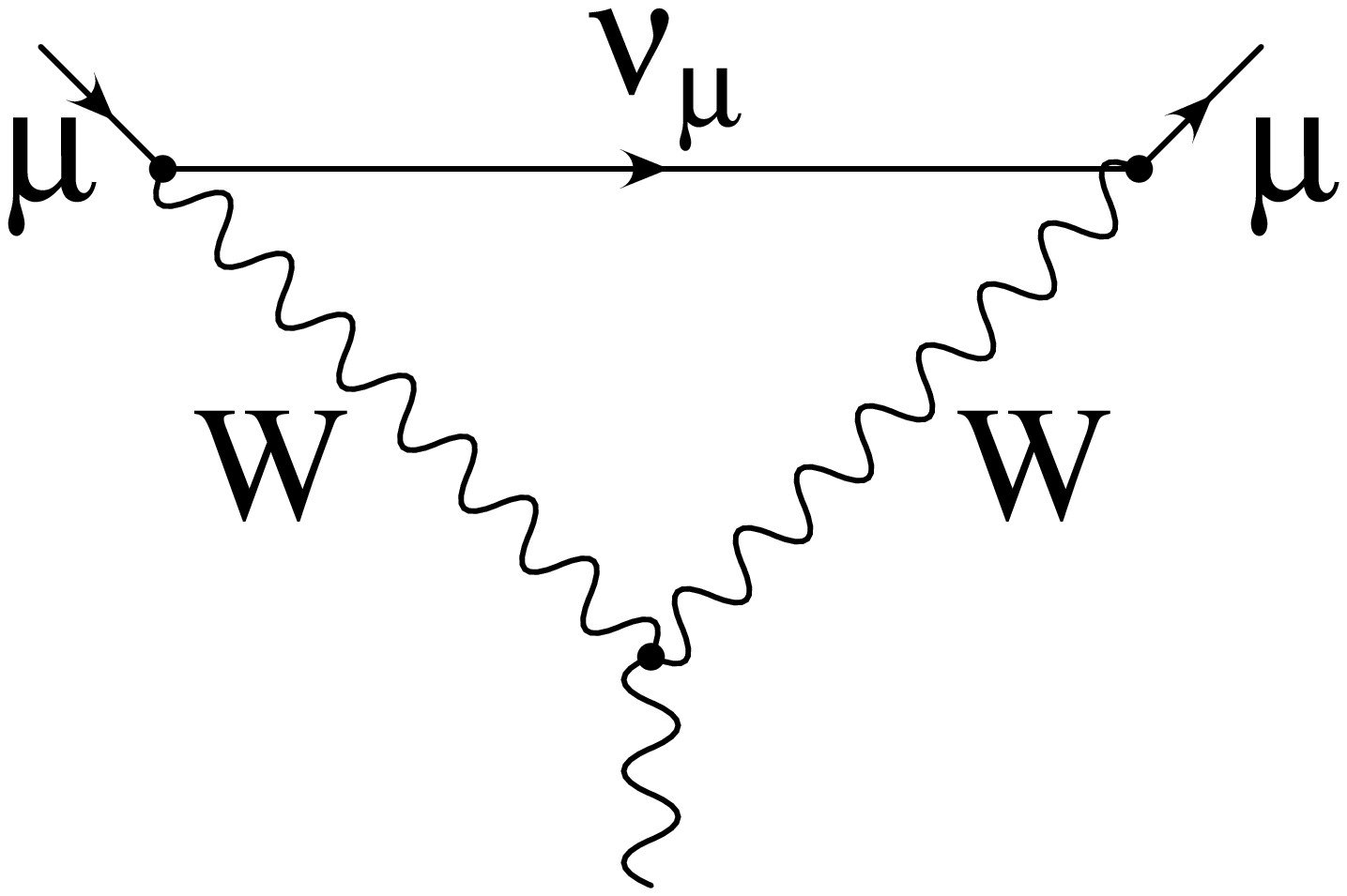,width=30mm,bbllx=210pt,bblly=410pt,bburx=630pt,bbury=550pt}
&\hspace*{.6cm}
\psfig{figure=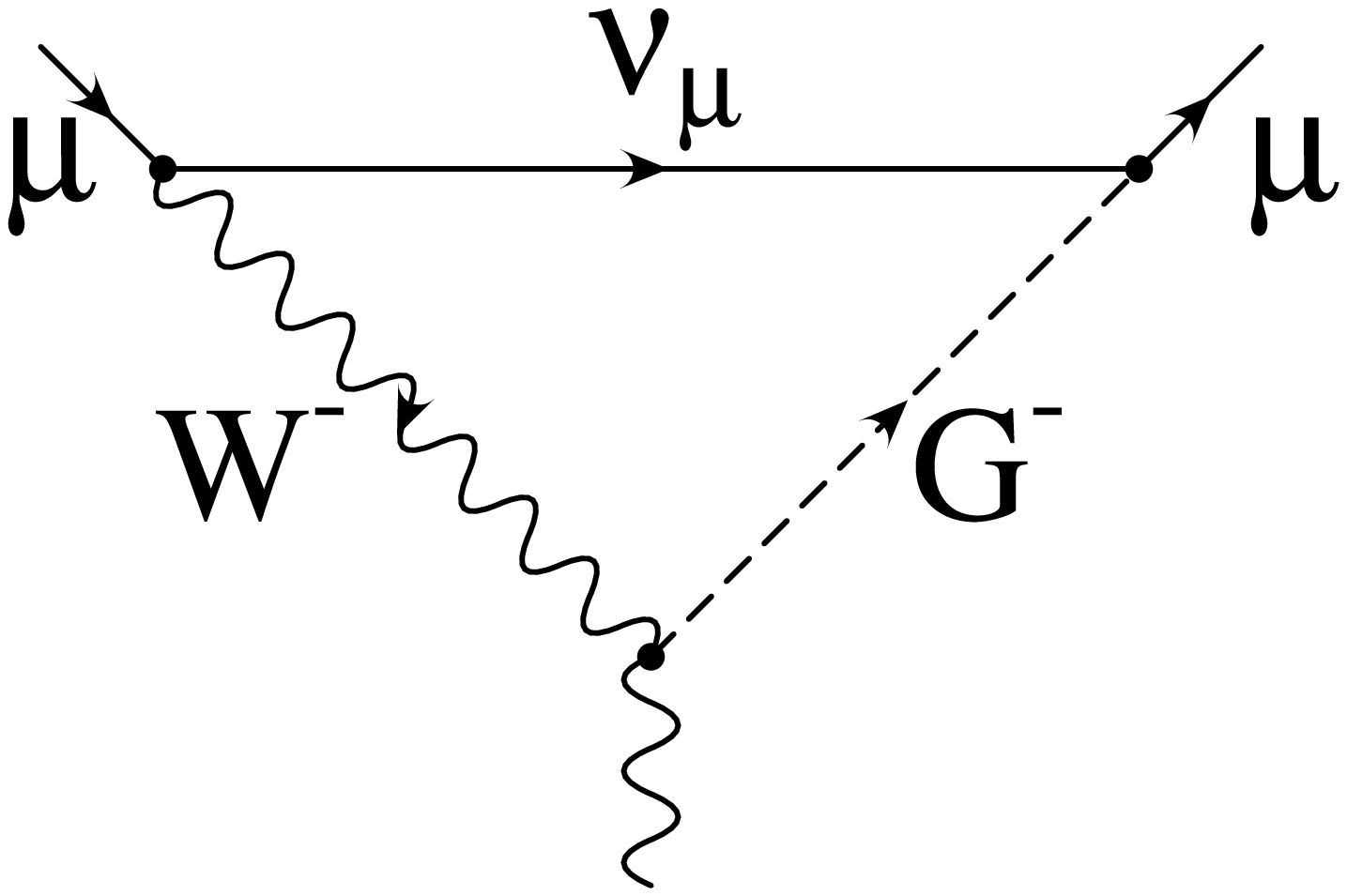,width=30mm,bbllx=210pt,bblly=410pt,bburx=630pt,bbury=550pt}
\\[5mm]
\rule{-18mm}{0mm} (a) &\hspace*{-.4cm} \rule{-6mm}{0mm}(b) & 
\hspace*{-.6cm}\rule{-2mm}{0mm}(c) 
\end{tabular}}
\]
\end{minipage}

\caption{One-loop electroweak corrections to $a_\mu$ (including the
QED contribution)} 
\end{figure}

\begin{figure}
\begin{minipage}{23.cm}
\hspace*{.8cm}
\[
\mbox{
\hspace*{-67mm}
\begin{tabular}{ccc}
\psfig{figure=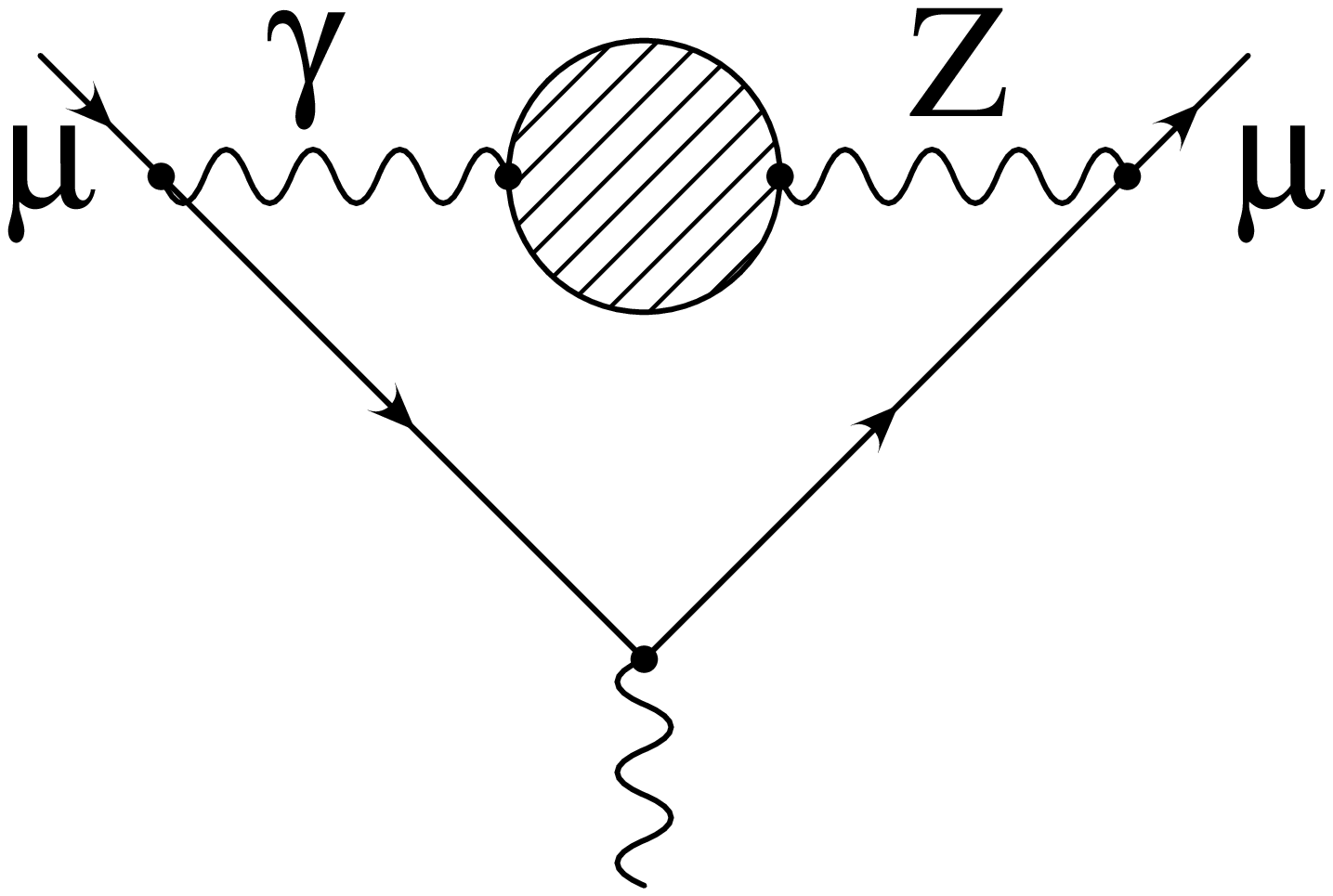,width=30mm,bbllx=210pt,bblly=410pt,bburx=630pt,bbury=550pt} 
&\hspace*{.6cm}
\psfig{figure=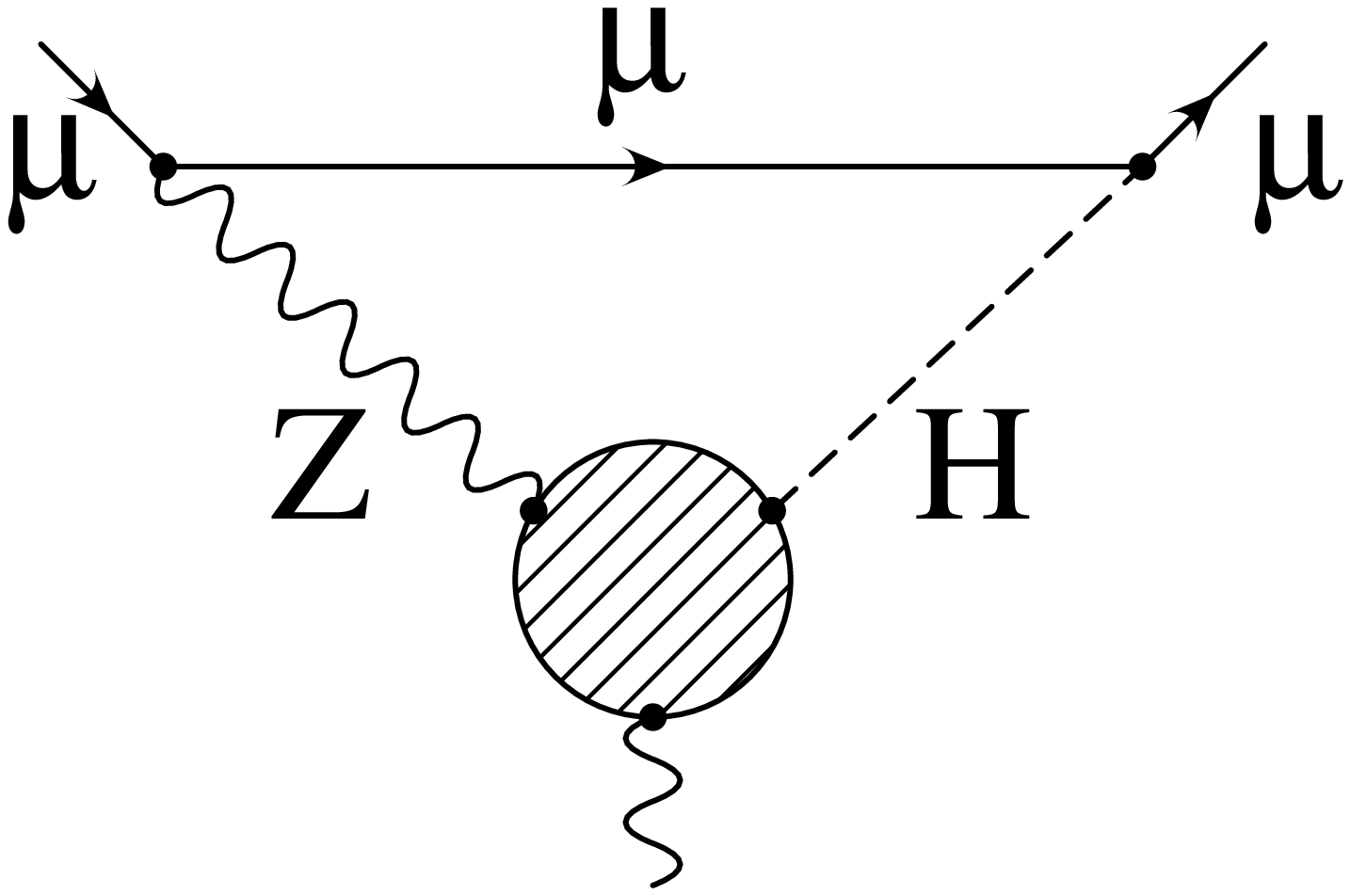,width=30mm,bbllx=210pt,bblly=410pt,bburx=630pt,bbury=550pt}
&\hspace*{.6cm}
\psfig{figure=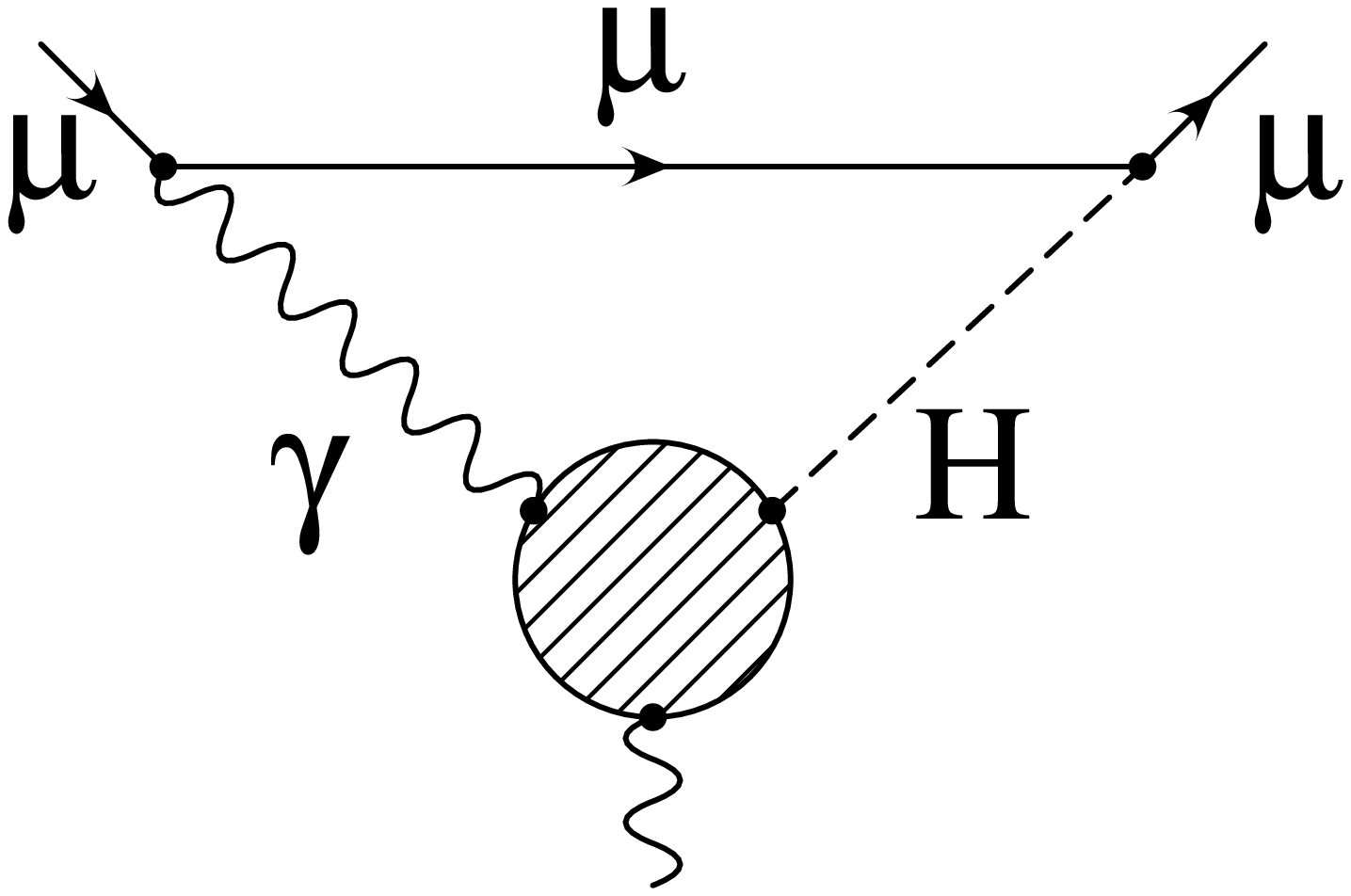,width=30mm,bbllx=210pt,bblly=410pt,bburx=630pt,bbury=550pt}
\\[5mm]
\rule{-17mm}{0mm} (a) &\hspace*{-.4cm} \rule{-6mm}{0mm}(b) & 
\hspace*{-.2cm}\rule{-6mm}{0mm}(c) 
\end{tabular}}
\]
\end{minipage}

\vspace{1cm}

\begin{minipage}{23.cm}
\hspace*{.8cm}
\[
\mbox{
\hspace*{-67mm}
\begin{tabular}{ccc}
\psfig{figure=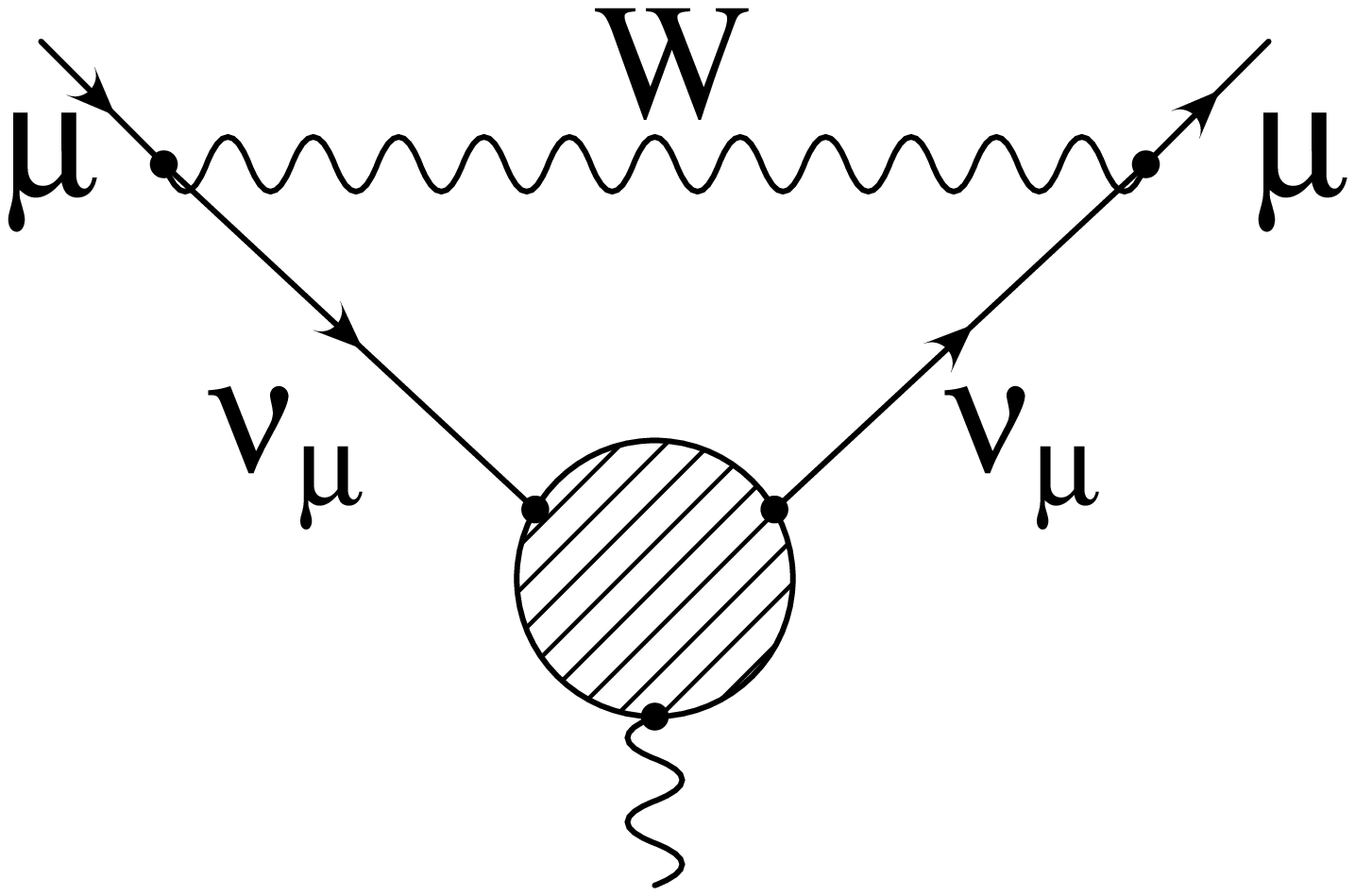,width=30mm,bbllx=210pt,bblly=410pt,bburx=630pt,bbury=550pt} 
&\hspace*{.6cm}
\psfig{figure=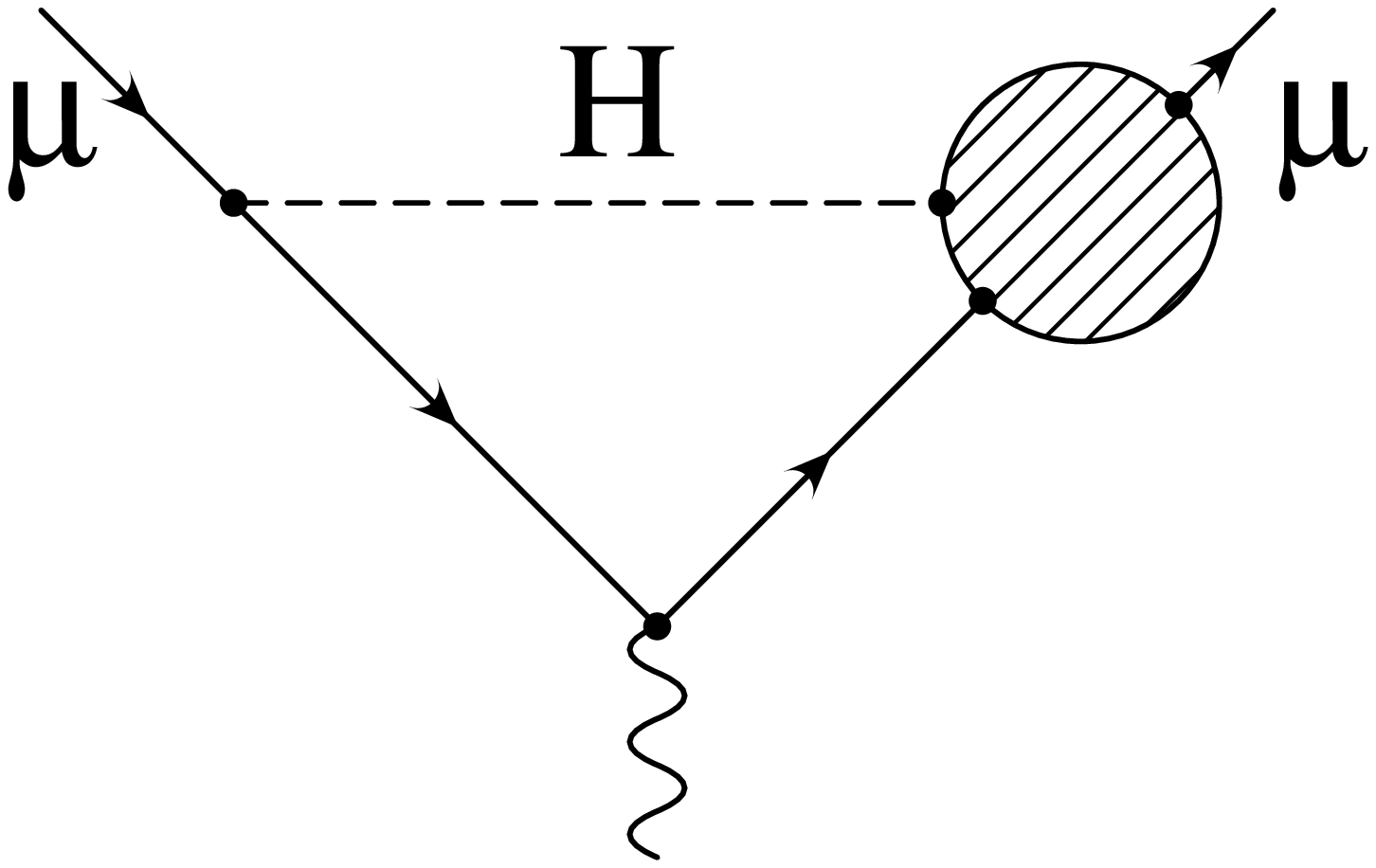,width=30mm,bbllx=210pt,bblly=410pt,bburx=630pt,bbury=550pt}
&\hspace*{.6cm}
\psfig{figure=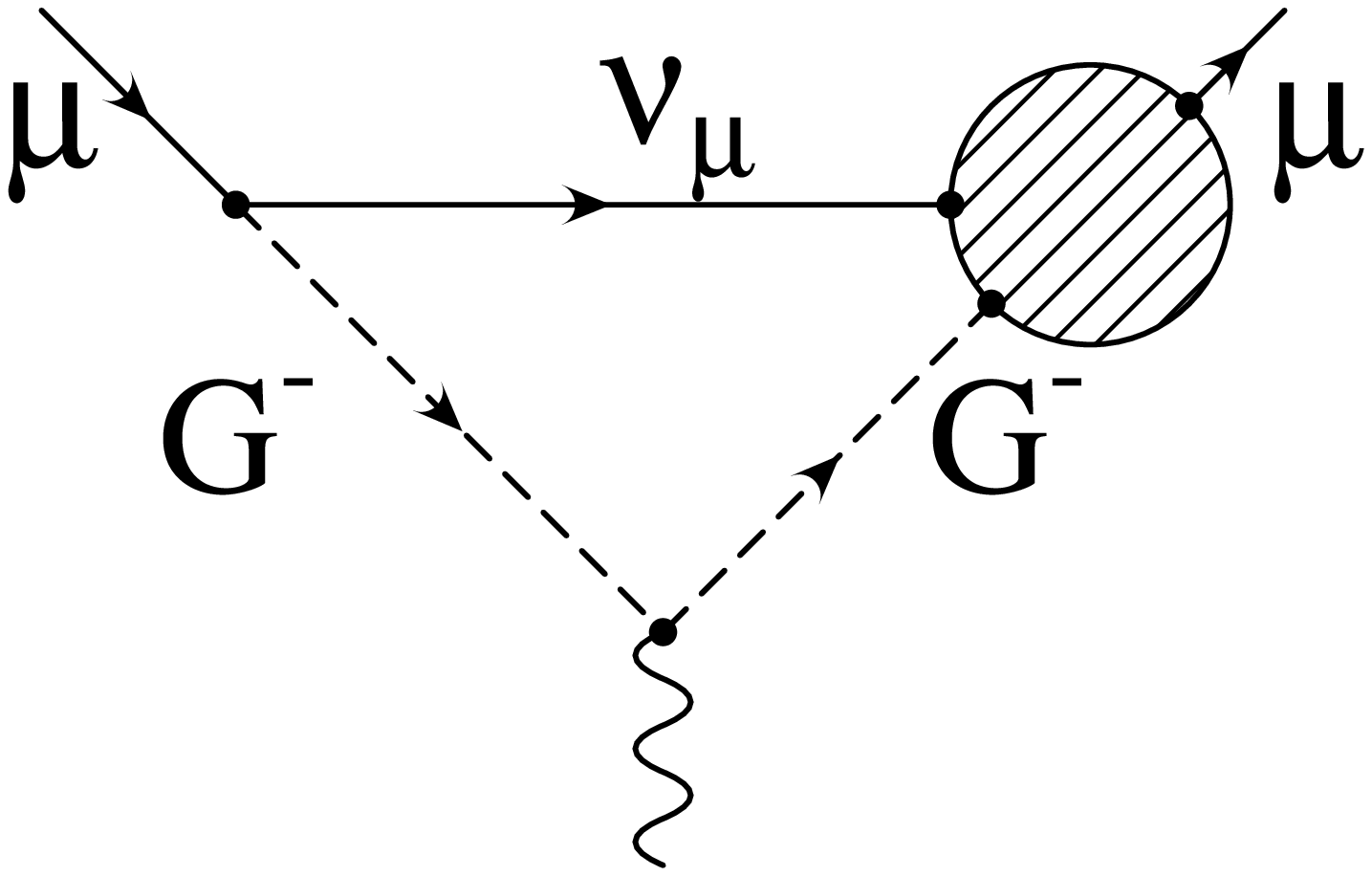,width=30mm,bbllx=210pt,bblly=410pt,bburx=630pt,bbury=550pt}
\\[5mm]
\rule{-17mm}{0mm} (d) &\hspace*{-.4cm} \rule{-6mm}{0mm}(e) & 
\hspace*{-.2cm}\rule{-6mm}{0mm}(f) 
\end{tabular}}
\]
\end{minipage}

\caption{New types of diagrams at the two-loop level}
\end{figure}

\begin{figure}
\vspace*{3mm}
\hspace*{6cm}
\psfig{figure=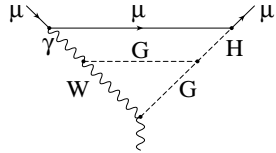,width=40mm,bbllx=10pt,bblly=410pt,bburx=630pt,bbury=550pt} 
\vspace*{9mm}

\caption{An effective $\gamma\gamma H$ coupling diagram which gives a
contribution to the leading logarithms in linear gauges.}
\end{figure}


\end{document}